\documentstyle[11pt,newpasp,twoside]{article}

\def\msun{M$_{\odot}$}
\def\rsun{R$_{\odot}$}
\def\fdg{\hbox{$.\!\!^\circ$}}
\reversemarginpar

\begin{document}
\title{Discovery of Eight Recycled Pulsars -- The Swinburne Intermediate Latitude Pulsar Survey}
\author{Russell T. Edwards}
\affil{Centre for Astrophysics \& Supercomputing, 
        Swinburne University of Technology
        Mail 31 PO Box 218 Hawthorn VIC 3122 Australia
}
\begin{abstract}
We have conducted a pulsar survey of intermediate Galactic latitudes
($5\deg < |b| < 15\deg$) at 20 cm.
The survey has been highly
successful, discovering 58 new pulsars, eight of which are recycled,
in only $\sim$14 days of integration time. One pulsar has a very
narrow ($2\deg$ FWHM) average profile for the pulsar's period (278 ms).
The six new recycled binary systems provide valuable information on the 
formation of white dwarf pulsar binaries. Two systems have massive white dwarf 
companions ($> 0.57$ \msun\ and $> 1.2$ \msun), while another
has a low mass ($\sim 0.2 $ \msun) companion in a 23.3-d orbit,
residing the well-known orbital period ``gap''.
\end{abstract}

\section{The Swinburne Intermediate Latitude Pulsar Survey}
Full details of the observing hardware and analysis procedures are
available in Edwards et al.\ (2000). Briefly, 265-s pointed
observations were made with the 64-m Parkes radiotelescope using the
sensitive new 21 cm 13-beam receiver.  The backend system includes
twenty-six filterbanks, each with ninety-six channels and a total
bandwidth of 288 MHz, centred at a sky frequency of 1374 MHz. Detected
filterbank outputs are summed in polarisation pairs and one-bit
digitised with an integration time of 125 $\mu$s.  Data is recorded on
magnetic tape for offline processing on the Swinburne supercluster, a
network of 64 Compaq Alpha workstations, using standard techniques
(e.g. Manchester et al.\ 1996).

The survey area has been observed and processed to a completeness of
90\% and has been highly successful with a minimal investment of telescope
time, discovering 58 new pulsars to date. Of these, eight are
recycled, six of which are in binary systems with circular orbits,
indicating white dwarf companions.

\section{Discovery Highlights}

We have discovered a pulsar with a period of 278 ms and an average
pulse profile only $W \simeq 2\fdg 1$ FWHM. Rankin (1990) observed
that the distribution of pulsar profile widths for core-type pulsars
is well fit by the constraint $W P^{1/2} > 2\fdg 45$, where $P$ is the
pulsar period in seconds.  The newly discovered pulsar
J1410--7407\footnote{Pulsar name is subject to change due to the
present uncertainty in its position}, however, has $W P^{1/2} =
1\fdg1$. Preliminary polarimetric results at 660 MHz and 1400 MHz
indicate that the profile has two components, furthering the
discrepancy between the observed component widths of this pulsar and
other long-period systems.

The orbital period distribution of low mass binary pulsars previously
appeared to include a ``gap'' (Camilo 1994) in the range $12.4 <
P_{\mathrm{orb}}\;\mathrm{(d)} < 56.3$. A number of authors (e.g.
Kulkarni 1995, Tauris 1996) have suggested that this gap separates
those systems that evolved with significant angular momentum losses
from those that did not.  Systems with an orbital period less than the
so-called ``bifurcation period'', $P_{\mathrm{bif}}$=1--2 days,
undergo orbital contraction during mass transfer due these losses.
The newly discovered binary system J1618--3919$^1$ has an orbital
period of 23.3 days, placing it in the middle of the ``gap''. Further
to the considerations above, we suggest that there is a narrow range
of initial orbital periods (significantly longer than
$P_{\mathrm{bif}}$) over which (for increasingly close orbits) angular
momentum losses quickly become significant. This results in an
under-density of systems with final orbits in the range of 7 -- 60
days, particularly around the upper end of this range (the former
``gap'').  The distribution around $P_{\mathrm{bif}}$ appears fairly
even.

The mass functions of two of the new recycled binary systems indicate
that the companion is a massive CO or ONeMg white dwarf -- for
J1757--5322 $M_{\mathrm{WD}} > 0.57$ \msun, whilst for J1157--5112
$M_{\mathrm{WD}} > 1.2$ \msun. Five recycled binary pulsars with
massive white dwarfs were previously known (see e.g. Arzoumanian,
Cordes \& Wasserman 1999). It has been suggested that the evolution of
these systems included a deep common envelope phase (van den Heuvel
1994). PSR J1757--5322 is in a very close 11-hour orbit (with an
orbital separation of $a \simeq 3.1$ \rsun) and may be explained by
this model only if an energy source other than orbital decay largely powers 
envelope ejection.
J1157--5112, on the other hand, is only marginally compatible
with the model of van den Heuvel (1994) due to the relatively
wide 3.5-day orbit ($a \simeq 13.5$ \rsun).

\acknowledgements
I thank the Galactic plane multibeam pulsar survey collaboration (see
Camilo et al.\ 1999) for their assistance of our use of equipment
built for that survey and helpful discussions, and Swinburne
collaborators M Bailes, M Britton and W van Straten.

\end{document}